\documentclass[letterpaper,10pt]{article}
\pdfoutput=1
\usepackage[italian,english]{babel}
\usepackage{amsmath}
\usepackage{amssymb}
\usepackage{braket}

\usepackage[a4paper,margin=2.5cm]{geometry}
\usepackage{graphicx}
\usepackage[leftcaption]{sidecap}
\usepackage{titling}

\usepackage[sort, numbers]{natbib}

\usepackage{booktabs}
\usepackage{xcolor}
\usepackage[font=small]{caption}
\usepackage{subfig}
\usepackage{hyperref}\hypersetup{
  colorlinks   = true, 
  urlcolor     = blue, 
  linkcolor    = black, 
  citecolor   = blue 
}

\begin{document}

\selectlanguage{english}
\thanksmarkseries{arabic}

\title{ \textbf{\Large Time-resolved multimode heterodyne detection\\ for dissecting coherent states of matter}}

\author{\normalsize Filippo Glerean$^{1,2}$, Giacomo Jarc$^{1,2}$, Alexandre Marciniak$^{1,2}$,\\ \normalsize Giorgia Sparapassi $^{1,2}$, Angela Montanaro $^{1,2}$, Enrico Maria Rigoni $^{1,2}$,\\ \normalsize  Francesca Giusti$^{1,2}$, Jonathan Owen Tollerud$^{1,2,3}$, and Daniele Fausti$^{1,2,4,*}$\\
\\
\small \emph{${}^1$Dipartimento di Fisica, Università degli Studi di Trieste, Via Valerio 2 Trieste I-34127, Italy}\\
\small \emph{${}^2$Sincrotrone Trieste S.C.p.A., Basovizza I-34127, Italy}\\
\small\emph{${}^3$Optical Sciences Centre, Swinburne University, 1 Alfred street Hawthorn, Victoria 3122, Australia}\\
\small\emph{ ${}^4$Department of Chemistry, Princeton University, Princeton, New Jersey 08544, USA}\\
\small  ${}^*$ daniele.fausti@elettra.eu \\
}

\date{April 7, 2020}
\maketitle

\pagenumbering{arabic}


\renewcommand{\abstractname}{} 
\begin{abstract}    
\normalsize
Unveiling and controlling the coherent evolution of low energy states is a key to attain light driven new functionalities of materials. Here we investigate the coherent evolution of non-equilibrium photon-phonon Raman interactions in the low photon number regime via femtosecond time-resolved multimode heterodyne detection. The time dependence of the weak probe spectral components is revealed by interferential amplification with shaped phase-locked reference fields that allow for a frequency selective amplification. We report measurements on $\alpha$-quartz that show that both amplitude and phase of the probe spectral components are modulated at the phonon frequency, but encode qualitatively different responses which are representative respectively of the time dependent position and momentum of the atoms. We stress that the sensitivity achieved here (1-10 photons per pulse) may be of relevance for both quantum information technologies and time domain studies on photosensitive materials. 
\end{abstract}

\section{Introduction}

The Raman response of materials and molecules is commonly measured in time domain through pump and probe (P$\&$P) experiments relying on Impulsive Stimulated Raman Scattering (ISRS). The most commonly employed scheme, dubbed one dimensional P$\&$P, uses two ultrashort light pulses: a first pulse (pump) excites the system while a second (probe), impinging on the sample at a controllable time delay, measures the time evolution of the non-equilibrium state. The measurement of the Raman response in P$\&$P experiments relies on the fact that impulsive photo-excitation (i.e. the interaction with light pulses on a timescale shorter than the period of the excitation in the material) triggers coherent non-equilibrium states of low energy excitations, such as phonons \cite{Hase1996,Sasaki2018}, magnons \cite{Kampfrath2010}, or electronic excitations \cite{Sagar_2007,Werdehauseneaap8652,Mansart4539}, whose time evolution can be subsequently characterized by the ultrashort probe pulse. ISRS has been historically introduced and discussed treating both the coherent vibrational state in matter, which is often dubbed coherent phonon, and the interacting electric field with a classical formalism \cite{Yan1985,Merlin1997,Stevens2002}.  More recently, ISRS has emerged as a powerful tool for quantum information and it has been shown that ISRS can be used to store/retrieve single photons in/from the elastic field of materials \cite{Lee1253,England2015} as well as frequency convert single photons through non-linear processes which can occur in probe pulses containing only a few photons \cite{Fisher2016,Glerean_2019}.  For this reason, there is a growing interest in studying coherent non-equilibrium Raman dynamics with low intensity pulses. 

In this paper, we report the first study of non-equilibrium coherent vibrational response in $\alpha$-quartz by means of probe light pulses with a few photons per pulse. We combine femtosecond spectroscopy with interferometric balanced heterodyne detection and pulse shaping to distinguish for the first time the spectrally selective amplitude and phase response imprinted on few photon probe pulses by the coherent phonon responses. We stress that with the methodology proposed the time and frequency resolution can be determined independently respectively by the pump and probe duration at the sample and the frequency resolution of the pulse shaper.

Optical heterodyne detection (OHD) is a linear interferential method in which the weak signal is mixed with, and amplified by, a strong  Local Oscillator (LO) field. The phase locked LO acts as an amplifier and as a phase reference that enable the separate measurement of the spectral amplitude and phase of the probe spectral components \cite{Benatti_2017,Esposito2015,Fittinghoff:96,Fuji_2000,Tokunaga:95,Lepetit:95,Tokunaga:96}. The signal to noise ratio of OHD can be improved by means of balanced detection \cite{Dorrer:01}. In fact, the balanced scheme removes the high intensity LO noise \cite{Carleton:68,Abbas:83} and significantly improves the detection sensitivity.

Moreover, differential detection, which directly subtracts the photocurrent prior to electrical amplification, reduces the electronic noise and vastly enlarge the detector dynamical range \cite{doi:10.1080/09500340.2013.797612}. 
Experiments studying spectrally resolved dynamics typically must tolerate high intensity probe pulses, providing an intensity in each spectral component high enough to be measured, and degraded signal to noise intrinsically limited by the dynamical range of multichannel detectors \cite{PhysRevB.93.054305,Novelli2014}. In this work, we have devised a technique to measure the spectrally resolved pump-probe signal which can operate with weak probe pulses in the low noise-floor regime provided by balanced detection. The spectral dependence of the signal is achieved by spectrally scanning a narrow LO across the probe spectrum.

The manuscript is organized as follows. We first describe how we employ a pulse shaper in order to perform phase stable heterodyne detection of the probe spectral components \cite{Forget:10} by selectively amplifying them through a shaped LO. Subsequently, we apply this methodology to study the time evolution of amplitude and phase of the different spectral components of the probe pulses interacting with a coherent vibrational response triggered by an ultrashort pump in $\alpha$-quartz, a benchmark material. Disentangling the spectral phase and amplitude time dependence allows us to distinguish between two different effects. The amplitude reveals the intra-pulse spectral density redistribution due to light-phonon energy exchanges which is proportional to the time evolution of the atom momentum, while the phase measures the linear modulation of the refractive properties, proportional to their displacement. Finally, we demonstrate that the dynamics of the spectral phase is a robust observable even in in measurements with probe pulses with few-photons.

\section{Multimode shaped balanced heterodyne detection}

\begin{figure}[h!]
    \centering
    \includegraphics[width=0.5\linewidth]{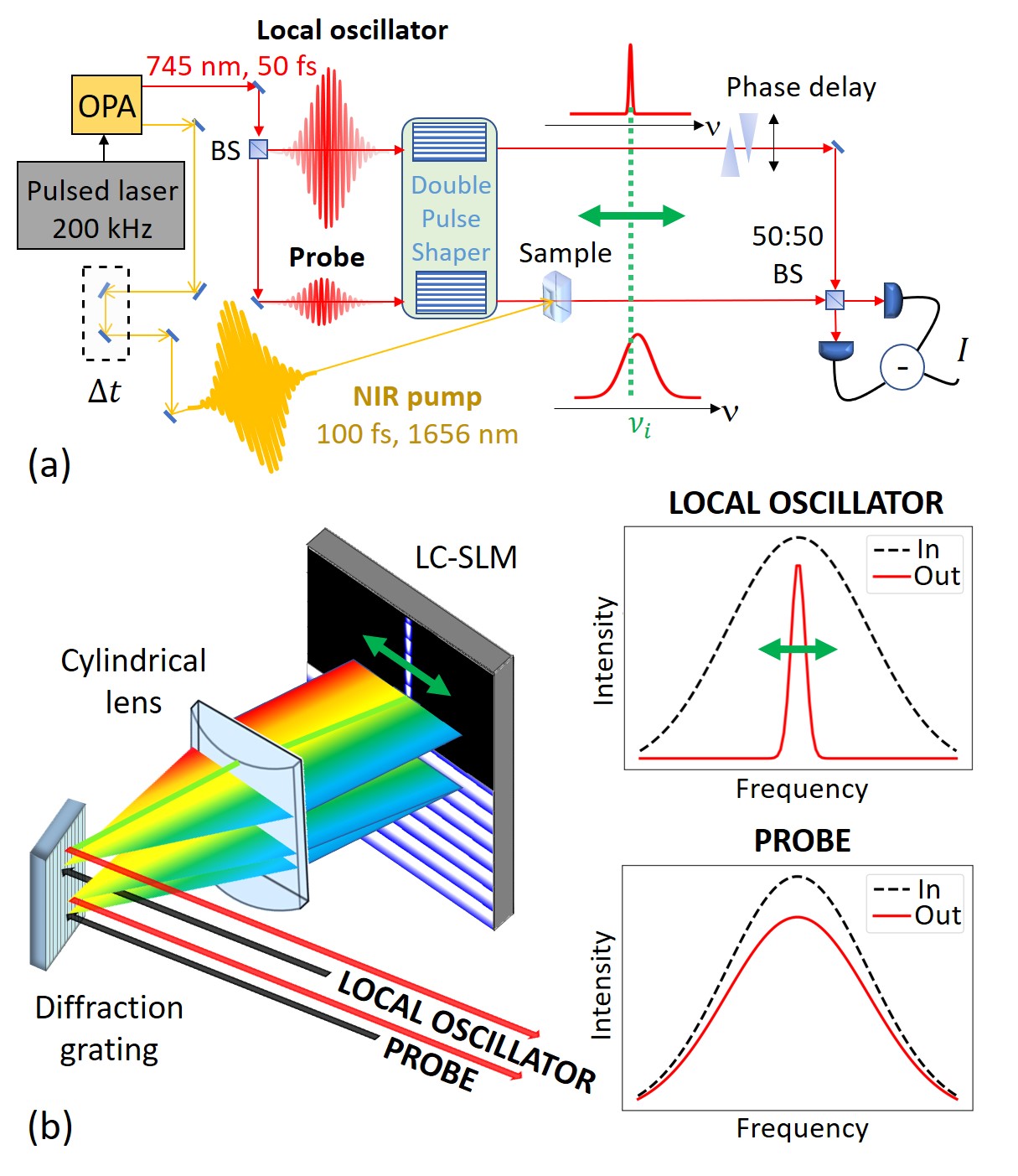}
    \caption{Experimental details. (a) A layout of the optical setup employed for time-resolved multimode heterodyne detection. It consists of a pump-probe experiment where the signal field out of the sample is amplified and referenced with a local oscillator shaped in its spectral content. (b) Two-beam diffraction-based pulse-shaper using a 2D phase only Spatial Light Modulator.}
    \label{fig:1}
\end{figure}

The experiment combines ultrafast pump-probe spectroscopy and multimode heterodyne detection of the probe field. The layout of the optical set-up is shown in Figure \ref{fig:1}(a) (polarization control and focusing elements are not shown). Pump and probe pulses are produced by a commercial 200 kHz pulsed laser+OPA system (Pharos + Orpheus-F, Light Conversion). The  signal from the OPA used as probe and heterodyning field in the interferometer, has a duration <50 fs and a tunable wavelength in the range 650-950 nm. The idler is used as pump and has a duration  <100 fs and wavelength in near infrared range. The experiments reported are performed with a probe (pump) wavelength  of 745 nm (1656 nm). Pump and probe both lie in the sample transparency range and have the same polarization.
The weak probe signal field interferes in a Mach-Zehnder scheme together with a more intense and phase-locked LO field, which acts as amplifier and phase reference. The outputs of the 50:50 beam splitter are acquired by two photodiodes and the difference between the two photo-voltages (heterodyne trace) is measured with a low noise charge amplifier and then digitized.  
It can be shown that the mean value of the differential current $\langle$I$\rangle$ as a function of the LO phase maps the mean phase-resolved quadrature of the probe field $\langle$X$\rangle$, which is the representative observable of the complex electric field. 

Indeed, by denoting with $z_\nu$ the amplitude of the single mode LO 
of frequency $\nu$, with $\phi_\nu^{LO}$ its tunable phase and with $\alpha_\nu=|\alpha_\nu | e^{i\phi_\nu }$ the complex amplitude of each probe mode, the heterodyne current reads:

\begin{eqnarray} \label{eq: 1}
\langle I \rangle &=& \sqrt{2}d\nu'|z_{\nu'}| \langle X_{\nu'}(\phi_{\nu'}-\phi_{\nu'}^{LO})\rangle \nonumber\\
&=&\int d\nu'|z_{\nu'}|(\alpha_{\nu'} e^{-i\phi_{\nu'}^{LO}}+ \alpha_{\nu'}^* e^{+i\phi_{\nu'}^{LO} })\\
&=&2\int d\nu'|\alpha_{\nu'}||z_{\nu'}|\cos(\phi_{\nu'}-\phi_{\nu'}^{LO}) \nonumber
\end{eqnarray}

In order to achieve frequency resolution, we select a narrow spectral band in the LO (approximated by a delta function, $z_{\nu'}=z_\nu \delta_{\nu,\nu'}$), which defines the heterodyne amplification of probe frequency $\nu$. Thus, the measured observable for each frequency mode is:
\begin{equation}
    \langle I \rangle_\nu = 2 |\alpha_{\nu}||z_{\nu}|\cos(\phi_{\nu}-\phi_{\nu}^{LO})
    \label{eq: 2}
\end{equation}

\begin{figure}[h!]
    \centering
    \includegraphics[width=0.5\linewidth]{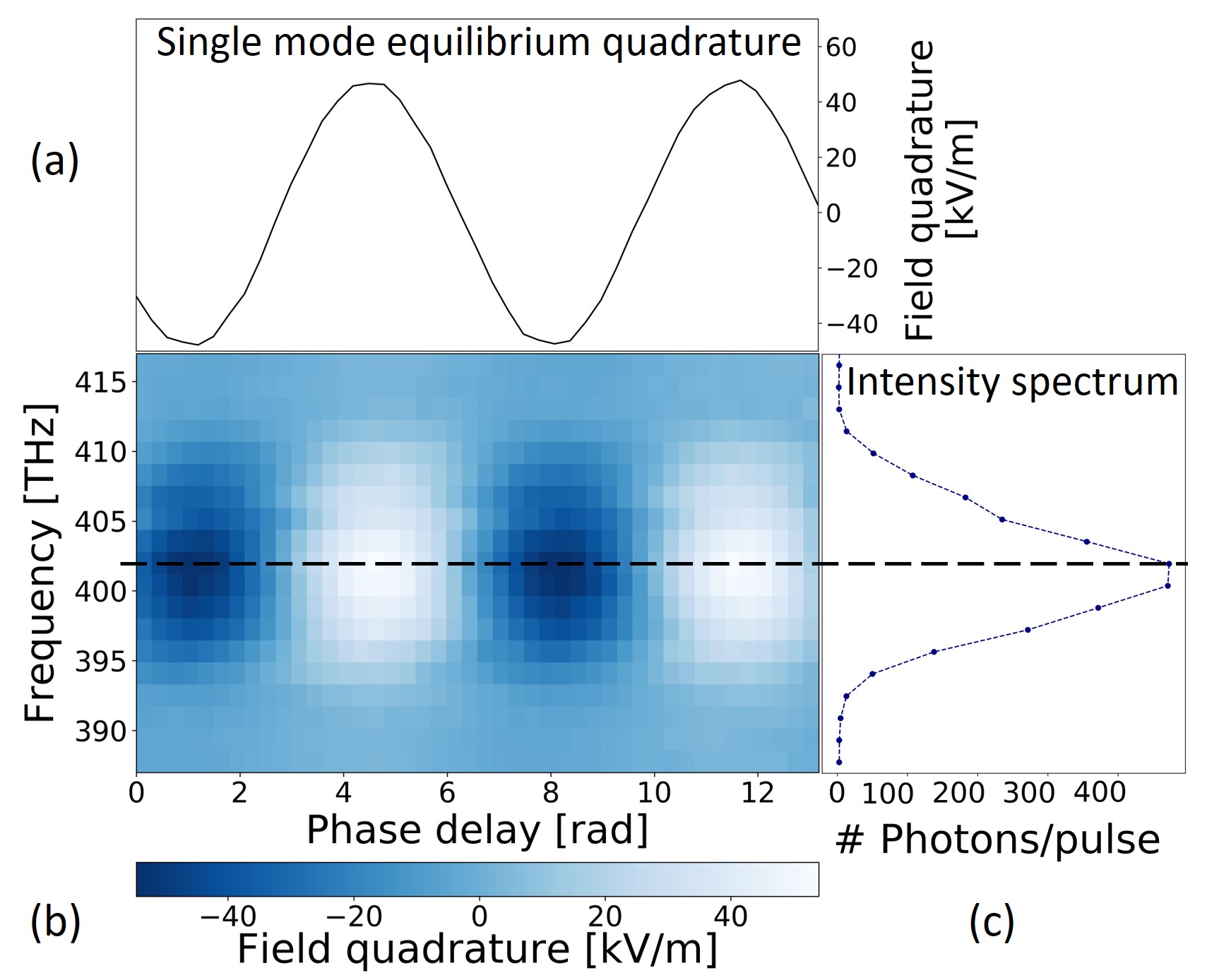}
    \caption{Equilibrium heterodyne spectral trace. a) Single mode quadrature in units of electric field at the sample focus. b) Spectral quadrature map. c) Corresponding intensity spectrum: we employ probe pulses with few hundreds of photons per selected mode.}
    \label{fig:2}
\end{figure}

The data recorded as a function of the relative phase between LO and signal consists in a sinusoidal oscillation at frequency $\nu$ (Fig. \ref{fig:2}(a)). The oscillation amplitude is proportional to the signal field amplitude, scaled by the LO field.

In the experiment, the LO narrow spectral band is selected by a diffraction based pulse shaper \cite{Weiner2000,Vaughan:05,Monmayrant_2010} (Fig. \ref{fig:1}(b)), which allows for an independent control of the amplitude and phase of each spectral component of the LO and the probe. The  horizontally dispersed spectral components are focused on a two dimensional Liquid Crystal-Spatial Light Modulator (LC-SLM) (Santec SLM-100) which consists of a multipixel array of nematic crystals used to generate programmable blazed diffraction gratings. The programmable vertical position and the  depth of the grating for all spectral components is used to control respectively the spectral phase and the amplitude of the diffracted beam. In particular, as depicted in Fig. \ref{fig:1}(b), applying a blazed pattern only to specific horizontal positions results in a narrowband diffracted LO.

In the experiment reported here, the pulse-shaper is used to narrow the LO to the width of a single mode, with a tunable central frequency and resolution of 0.1THz. As illustrated in Figure \ref{fig:1}(b), the probe is also shaped. The choice of common optics between LO and probe improves the interferometer stability, and the capability of the LC-SLM to modulate the phase of each spectral component can be exploited to correct the probe temporal compression and its chirp (see supplementary material for details on the setup and characterization).
The reconstruction of the quadrature on the entire spectral range is performed by scanning the LO across the different frequency components. For each selected mode, we measure the heterodyne oscillation trace (Eq.\ref{eq: 2}, Fig. \ref{fig:2}(a)) and by scanning  the LO frequencies we obtain the spectral map in Figure \ref{fig:2}(b). Taking into account that the employed LO has the same spectral shape of the probe, i.e. $z_\nu \propto \alpha_\nu$, we can renormalize the heterodyne oscillation and get the probe amplitude spectrum. By squaring it we can obtain the intensity of each spectral component ($I=| \alpha |^2$) in Figure \ref{fig:2}(c).  In order to perform a quantitative estimation, we record the intensity of the signal and LO beams with a power meter.  We employ a 402 THz (745 nm) beam, 20 THz bandwidth. The probe has 5 fJ per pulse ($10^4$ photons). The LO is three orders of magnitude more intense: 5 pJ, $10^7$ photons. The single modes are obtained with a 0.25 THz bandwidth filter and for the central LO mode we have 0.12 pJ, $10^6$ photons. All this considered, acquiring 2000 pulse repetitions for each LO phase point and sampling the quadrature oscillation with 0.25 rad steps, we can reconstruct the mean value intensity spectrum with few hundreds of photons per mode, with an uncertainty of $\pm 3$ photons.

\section{Time-resolved coherent vibrational dynamics in quartz}

\begin{figure}[h!]
    \centering
    \includegraphics[width=0.5\linewidth]{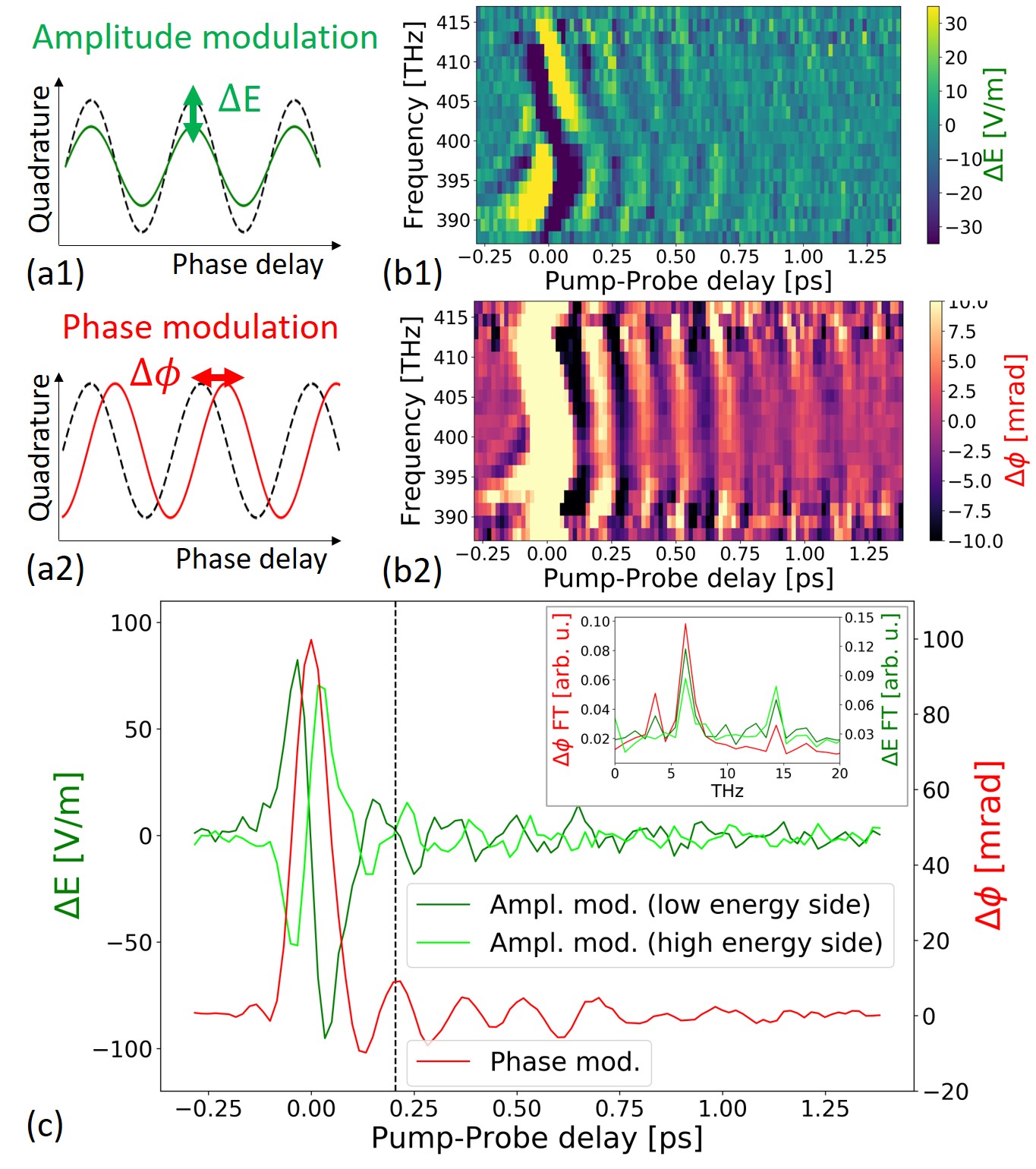}
    \caption{Time-resolved vibrational dynamics. 
a) Cartoon of non-equilibrium amplitude (a1) and phase (a2) quadrature modulations. b) Spectral maps of coherent phonon oscillations in quartz for amplitude (b1) and phase (b2). c) The amplitude response has opposite behavior on the two sides of the spectrum, while the phase response is the same across the spectrum and is shifted by $\pi$/2 (see dashed vertical guide). Fourier Transform analysis of the time traces (insert) shows the detected phonons frequencies; considering the background level, phase detection grants a better signal to noise ratio compared with amplitude detection.}
    \label{fig:3}
\end{figure}

We apply multimode heterodyne detection in time domain to address the coherent evolution of lattice vibrations in non-absorbing media by studying how the probe-phonon interaction is mapped into the spectral modifications (amplitude and phase) of each probe frequency component (Fig. \ref{fig:3}(a)). Transparent media (as $\alpha$-quartz) represent benchmark systems for this study, since no electronic transitions are dipole allowed within the pump or probe bandwidth and the probe-matter interaction can be treated as an effective photon-phonon coupling \cite{Yan1985,Merlin1997,Stevens2002,Glerean_2019,Benatti_2017,Esposito2015}. 

The energy transfer in photon-phonon interaction is mediated by non-resonant Impulsive Stimulated Raman Scattering (ISRS) \cite{Yan1985,Glerean_2019,Benatti_2017,Esposito2015} which is an intrinsic multimode process occurring whenever an ultrashort pulse (with a bandwidth greater than the excited phonon frequency $\Omega$) propagates in a Raman-active medium. All optical modes in the optical pulse with an energy difference corresponding to $\hslash \Omega$ can interact and create (Stokes Raman process) or annihilate (Anti-Stokes) a phonon.

The  coherent vibrational states are excited via ISRS by the intense pump pulse (fluence 10 mJ/cm$^2$). The time dependent modulations of spectral amplitude and phase of the weak probe (50 pJ/cm$^2$) are presented in Figure \ref{fig:3}(b). The latter are obtained by separately fitting the pumped and equilibrium heterodyne current with a sinusoidal function for combination of delay and frequency, and then we evaluate the differential amplitude and phase from the fitted parameters. We use a chopper blade on the pump beam to simultaneously reference the fluctuations in the equilibrium response and, in this way, we can observe spectral variations with a sensitivity of 2 photons per pulse, and a phase stability of 2 mrad (details of noise analysis can be found in the supplementary data).

We observe an oscillation in time of the response of both amplitude and phase at the frequency of the excited phonon modes. Interestingly, the amplitude and phase responses are out of phase and reveal  two different responses, representative of two typologies of interaction, ruled by the coherent phonon phase\cite{Glerean_2019}. In particular,  as highlighted in Fig. \ref{fig:3}(c) the amplitude is modulated with opposite sign on the two sides of the probe spectral bandwidth. This indicates a dynamical spectral weight redistribution within the probe bandwidth through ISRS. If the probe interacts with the coherent vibrational state at a time when the atoms have the maximum momentum the Stokes process dominates and high frequency photons are down-converted. On the contrary, when the probe impinges on the sample when the atoms are coherently moving with minimum momentum a partial quench of the atomic motion can be triggered and the probe-phonon interaction is dominated by the Anti-Stokes process; i.e. low frequency photons are up-converted and an effective energy transfer between the elastic field and the probe occurs.  
Conversely, the time dependence of the phase of the different spectral component exhibit a uniform behavior across the probe spectrum, which results from the linear modulation of the sample refractive properties. The refractive index is dependent on the lattice displacements and the  time domain response of the phase is ruled by the coherent phonon position coordinate. In Figure \ref{fig:3}(c), we verify the dependence on phonon phase space operators – position and momentum – by noting that amplitude and phase dynamics are $\pi$/2 shifted. 

Quantitatively, in this experiment we have a few hundreds of photons per mode per pulse and 1-10 photons modulation. Looking at the phonon spectra obtained through Fourier transform analysis (insert in Fig. \ref{fig:3}(c)), we highlight that the signal-to-noise ratio for the phase dynamics is better than the amplitude one. This can be understood by considering that the pump-induced amplitude modulation is due to a small fraction of scattered photons, while the phase modulation is the result of a process which involves the entire probe beam. This suggests that, in the low probe photon number, phase modulation is a more suitable observable compared with amplitude or intensity dynamics.

\section{Conclusion}

In this paper, we present time-resolved multimode heterodyne detection and measure non-equilibrium dynamics associated with a coherent vibrational response in a prototypical sample ($\alpha$-quartz) in the probe low photon number regime.

As a perspective, the fact that quantitatively, we work with few hundreds of photons per frequency mode, means we can access the shot noise regime, where the probe photon statistics is dominated by quantum fluctuations, which overcome the classical instabilities (which scale linearly on the order of 0.1-1\% for our laser source). 

Reaching the shot noise limited regime enables to perform full quantum state reconstruction of the optical field and to merge the analysis of statistical degrees of freedom of light with ultrafast non-equilibrium processes \cite{Grosse:14}. In particular, ISRS is an intrinsically multimode process which imparts correlation among the different spectral components coupled through the phonon modes \cite{Tollerud5383} and could represent a new platform to build and manipulate multimode entanglement within the bandwidth in optical pulses \cite{Roslund2013,Cai2017}.

\section*{Acknowledgments}
This work was supported by the European Research Council through the project INCEPT (grant Agreement No. 677488). 


\newpage
\appendix

\setcounter{figure}{0}
\renewcommand{\thefigure}{S\arabic{figure}}

\begin{center}
\textbf{\Large Supplementary material\\ }
\end{center}

\section{Data analysis and noise estimation details.}

The observable we are considering in our heterodyne detection scheme is the quadrature of the field. A single quadrature trace is recorded acquiring a train of multiple pulses for each considered quadrature phase (i.e. Local Oscillator (LO) delay).
The time-resolved dynamical response induced at fixed delay between pump and probe is obtained by comparison between the quadrature of the perturbed (pumped) and reference (unpumped) datasets. The train of pulses (2000 reps at 200 kHz) is sorted between the two sets by means a chopper blade (400 Hz) placed on the pump beam. 
The quadrature obtained with each dataset as a function of selected frequency mode $\nu$ and pump delay $\Delta$t is fitted with the function 

\begin{equation}
\tag{S1}
    I(\nu,\Delta t)_{p/u}^{fit} = k_\nu E_{p/u} sin(2\pi \nu + \phi_{p/u}) + \delta_{p/u}
\end{equation}

where the label p/u indicates respectively pumped or unpumped sets. 
The points in the spectral field amplitude and phase modulations maps presented in the main are obtained from these fit parameters as: 

\begin{equation}
\tag{S2}
    \Delta \phi (\nu,\Delta t) = \phi_p - \phi_u
\end{equation}

and

\begin{equation}
\tag{S3}
    \Delta E (\nu,\Delta t) = E_p - E_u
\end{equation}

\begin{figure}[h!]
    \centering
    \includegraphics[width=16cm]{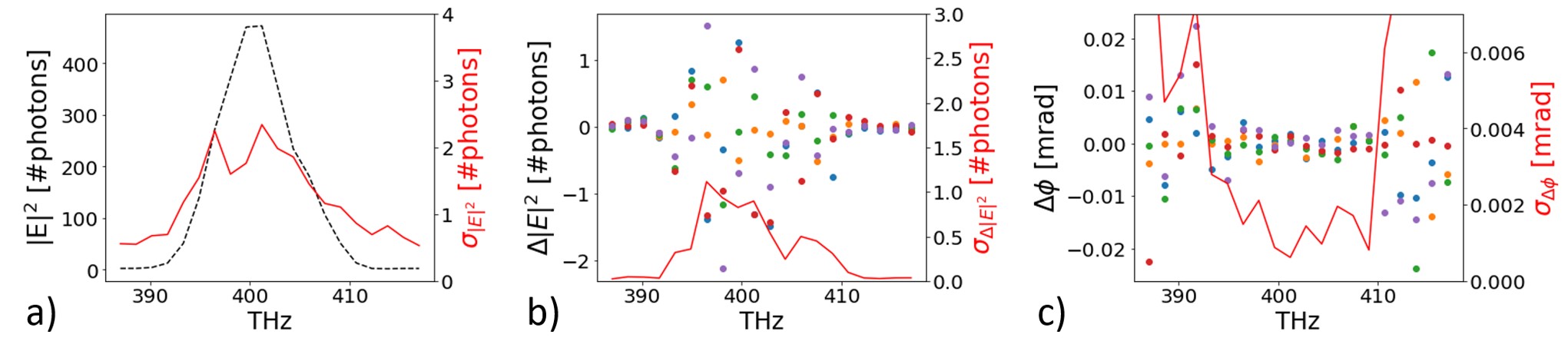}
    \caption{Quantitative results of photon number and sensitivity estimations.
(a) Intensity spectrum of the probe quadrature (dashed) and relative standard deviation (solid) in units of estimated number of photons. 
(b) Differential intensity modulation sensitivity in photon units. We plot the repetitions (dots) measured at equilibrium and their relative standard deviation (line), which results less than 2 photons per pulse.
(c) Differential phase modulation sensitivity. The interferometric stability on the acquisition timescale is calculated as standard deviation (line) between repeated equilibrium acquisitions (dots) and, in the probe frequency range, is better than 2 mrad.
}
    \label{fig:s1}
\end{figure}

Regarding the frequency dependent parameter $k_\nu$ accounts for the LO intensity and the expression in the desired units of measurement.
In order to estimate the probe pulse intensity, $|E_u |^2$, in terms of number of photons (Fig. \ref{fig:s1}(a)), we record probe and LO powers. In particular, we measure the weak probe beam power prior to attenuation with a filter of known specs. Then, we derive the photon number estimation and accordingly renormalize the spectrum obtained with the quadrature measurement. The associated uncertainty is calculated propagating the standard deviation on the values obtained with repeated quadrature acquisitions (i.e. multiple scans of the LO phase delay). In the examined configuration, we obtain an intensity spectrum of few hundreds of photon per mode (0.25 THz shaping bandwidth) with an accuracy better than 3 photons. 
To evaluate our sensitivity to time-resolved modulations, we repeatedly measure the differential signal between the chopper sorted datasets without pump beam. This way, by considering the standard deviation on multiple quadrature acquisitions, we are monitoring the equilibrium noise level. The referencing at the chopper frequency allows us to compensate for fluctuations on timescales slower than 400 Hz.
We report the differential signal data at equilibrium (i.e. no pump) for spectral intensity (Fig. \ref{fig:s1}(b)) and phase (Fig. \ref{fig:s1}(c)). As expected, the data are distributed around zero modulation (equilibrium). Thanks to the chopper referencing, we note that the uncertainty on the intensity modulation is lower than the one on the absolute value and results in less than 2 photons per pulse, per mode.
Regarding the phase, we are able to distinguish non-equilibrium modifications as small as 2 mrad.

\section{Capabilities for spectral phase shaping and retrieval.}

In the main text, we highlight the role of pulse shaping as a method to independently tune and select the amplitude of each spectral component. Besides this, it also allows the control of the spectral phase in a frequency-resolved way. Taking this into account, we show that implementing the Spatial Light Modulator (SLM) in our heterodyne detection scheme makes our setup capable of both shaping and retrieving the spectral phase of the probe pulse, relatively to the Local Oscillator (LO) one. 
Since heterodyne interference is a linear process, it is not possible to directly infer the pulse spectral phase. However, we can obtain this information with respect to a reference, which in this case is the LO. Thus, by characterizing the latter, we can in principle reconstruct the spectral phase of weak pulses, whose response in non-linear processes could not be detected.

\begin{figure}[h!]
    \centering
    \includegraphics[width=0.6\linewidth]{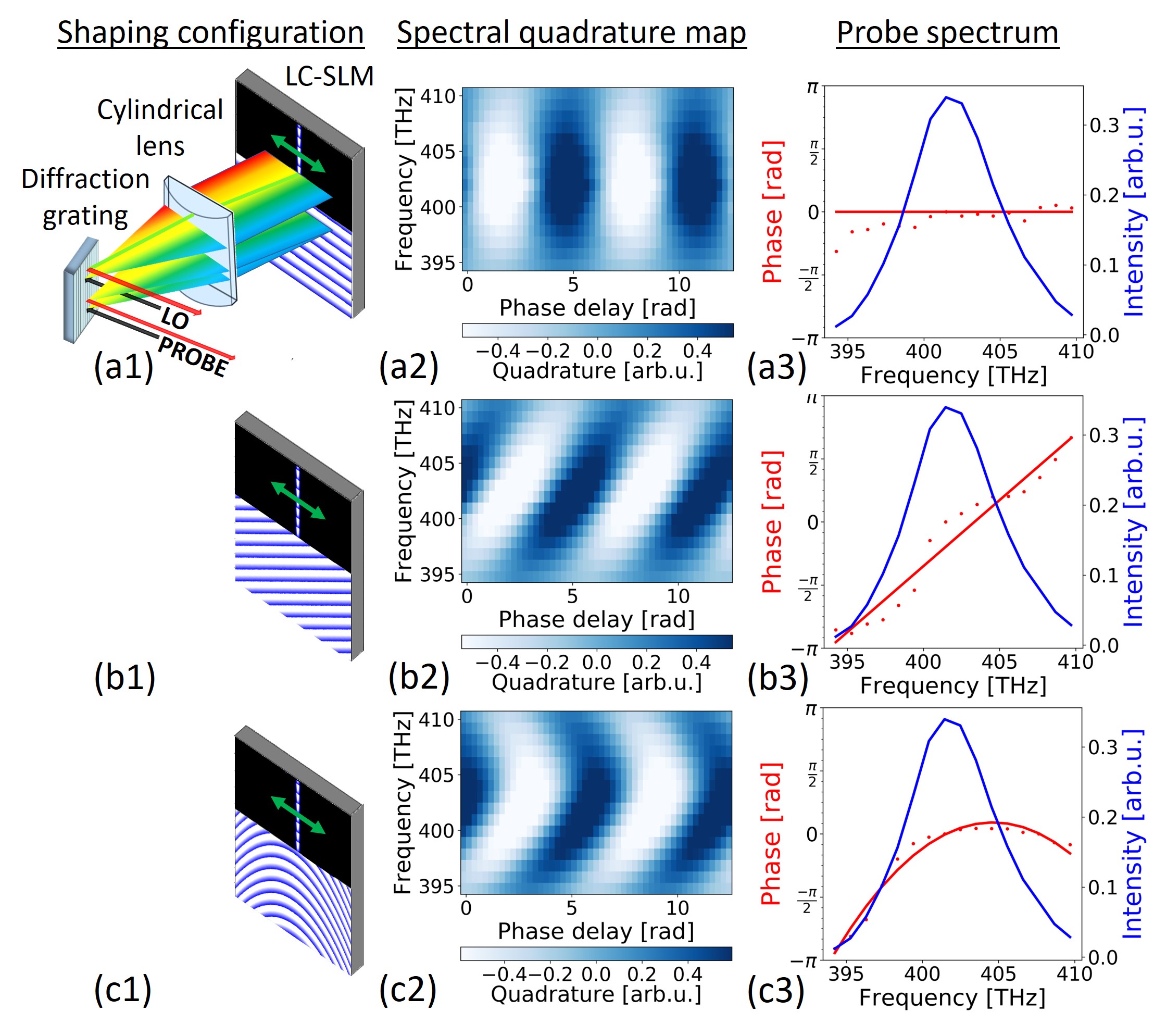}
    \caption{Spectral phase shaping of static quadrature by mean of the LC-SLM and corresponding retrieval of the probe phase and intensity spectrum through multimode heterodyne detection. 
In the first column (1) we present the programmable grating pattern applied with the SLM. The measured frequency resolved quadrature is reported in (2) and the relative spectral intensity (blue) and phase (red, data points and fit) are plotted in (3). We test our approach for three case studies:
(a) Spectral phase of the unshaped probe.
(b) Linear spectral phase modulation of the probe pulse. It corresponds to imparting a time delay to the probe pulse.
(c) Quadratic spectral phase modulation as a function of the probe frequency. It corresponds to imparting a temporal linear chirp to the probe pulse, hence enabling to control its temporal compression.}
    \label{fig:s2}
\end{figure}

In this appendix, we present a couple of examples two illustrate the potentiality of this method. Firstly, we illustrate how the SLM programmable gratings are employed to control and modify the spectral phase. Then, we apply the frequency resolved heterodyne detection described in the main as a tool to retrieve the imprinted feature.
The starting point is a configuration where both probe and LO pulses are Fourier Transform limited. If we leave the phase structure unperturbed, which corresponds to a flat phase of the grating pattern (Fig. \ref{fig:s2}(a1)), by measuring the quadrature (Fig. \ref{fig:s2}(a2)) map we obtain a constant quadrature phase, which correspond to zero relative spectral phase difference between probe and LO.
As a first example of modulation, we set a linear phase pattern in the SLM grating (Fig. \ref{fig:s2}(b1)). We observe it results in a linearly evolving phase in the quadrature map (Fig. \ref{fig:s2}(b2)). This way we retrieve the linear spectral phase pattern set by the shaper (Fig. \ref{fig:s2}(b3)), which from the physical point of view is imparting a temporal shift to the probe pulse.
Furthermore, we report how we are also able to control the pulse compression (i.e. the chirp) by applying a parabolic pattern (Fig. \ref{fig:s2}(c1)), and reconstructing in this case a quadratic spectral phase (Fig. \ref{fig:s2}(c2,3)).

\end{document}